# ENERGY DEPENDENT PROMPT NEUTRON MULTIPLICITY PARAMETERIZATION FOR ACTINIDE PHOTOFISSION

A.I. Lengyel[1], O.O. Parlag[1], I.V. Pylypchynec[1], V.T. Maslyuk[1],
M.I. Romanyuk[1], O.O. Gritzay[2]

*[1] Institute of Electron Physics*
*Universitetska 21, Uzhgorod 88017 Ukraine, myk.romanyuk@iep.org.ua*
*[2] Institute for Nuclear Research*
*Prospekt Nauky 47, Kyiv 03028 Ukraine*

The prompt neutron averaged number $\bar{\nu}$ and the dependence of prompt neutron yield on fragment mass $v(A)$ of photofission of actinide nuclei $^{235}$U and $^{238}$U in the giant dipole resonance energy range has been parameterized. This allows us to describe the observed changes of saw-tooth behavior of neutron yield from the light and heavy fragments using few energy and nucleon composition dependent free parameters and to predict $v(A)$ for other actinide isotopes.

## 1. INTRODUCTION

The number of prompt neutrons is important nuclear-physical parameter necessary for practical calculations. This value is determined in detail and accurately for neutron-induced reactions for the most nuclides. At the same time, the experimental data and evaluation of prompt neutron yield in the case of photofission are much scarcer. With the increasing interest in the methods of nuclear fuel burning and long-lived actinide decontamination the need in precise values of nuclear constants is evident during last years [1-6]. This is especially true for photonuclear constants. Therefore, the search was focused on a formulas that can be used to estimate the value of average number $\bar{\nu}$ $(A_F)$, $A_F$ – mass of fissioning nuclide, and number of neutrons $v(A)$, emitted by corresponding fission fragments of atomic mass for photofission of arbitrary actinide [7].

The same data are also used to obtain the mass and charge distribution of actinide nuclei fission products and to convert the secondary fragments (products) yields into the primary ones. Typically, to estimate the number of neutrons $v(A)$, emitted by corresponding fission fragments of atomic mass $A$, the phenomenological Wahl method [8] is applied. This method is widely used till present for estimation of $v(A)$ during neutron- and gamma-induced fission [9-11]. However, it does not reflect the complex structure of sawtooth-like $v(A)$ dependence due to nuclear shells effect.

So far there is only information about the average number of neutrons emitted by two conjugated fragments and there is no information on photofission neutron emission curves. Primarily, this is due to experimental difficulties of time-span or direct neutron measurements in photofission experiments. However, in principle, neutron emission curves can be obtained combining the measurement of fragments mass distribution and post-fission neutrons.

Calculations of ν(A) for $^{238}$U photofission based on the parameters of asymmetric fragments mass distribution and the average total neutron yield $\bar{\nu}(A_F)$ provide only qualitative, approximate reflection of saw-tooth behavior [12].

There is, however, another, more advanced method to determine the neutron emission curves ν(A), based only on the mass distribution of fission fragments and products using the so-called Terrell technique [13]. This method was used to calculate prompt neutron yields ν(A) at photofission of $^{232}$Th, $^{235}$U and $^{238}$U [14,15], based on technique [13,16] and data from the primary [17,18] and secondary [19,20] mass distributions of photofission fragments and by seven-mass-point approach for wide range of neutron induced fission nuclides [21].

The question is whether it is possible to identify some of the basic laws using the results of calculations depending on such parameters of actinides photofission as charge, mass, photon energy, shell characteristics. According to the results, one can parameterize neutron yields depending on the mass of actinide photofission fragments and other parameters that would reproduce the complex structure of saw-tooth behavior and allow predicting the dependence of ν(A) for the photofission of a wide class of actinide nuclei. This is a task of extreme interest [22].

## 2. PARAMETERIZATION OF ν(A) AND RESULT OF CALCULATION

A recent analysis [23, 24] of experimental data on neutron yields from fragments of thermal neutron fission of $^{233}$U, $^{235}$U, $^{239}$Pu and spontaneous fission of $^{252}$Cf showed that for a detailed account of "saw-tooth" particularity of dependence of fission neutron yield from a mass, an efficient tool is the value of R(A), introduced by Wahl [25], which is defined as

$$R(A) = \nu_{L,H}(A)/\bar{\nu}(A), \tag{1}$$

where $\nu_{L,H}(A)$ - prompt neutron yield of light and heavy fragment mass respectively, $\bar{\nu}(A)$ - total neutron yield, A - fragment mass and consists of several segments to reflect the observed features, depending on the complexity of the experimental behavior of R(A). Therefore, the whole range of fragments mass was divided for more then 2 x 4 segments [23,26]).

"Experimental" values of $R(A)$ (with errors) can be determined using formula (1) from experimental values of $v_{L,H}(A)$ and $\bar{v}(A)$.

Model function $R(A)$ is chosen as a linear function for each segment:

$$R_i^L(A) = a_i^L + b_i^L(A - A_L), \qquad (2)$$

$$R_i^H(A) = 1 - R_i^L(A - A_H) \qquad (3)$$

for light and heavy fragments, respectively, $i$ - number of the segment, $a_i^L$, $b_i^L$ $A_L$ - parameters, $A_H = A_f - A_L$, $A_f$ - mass of compound nucleus.

The total number of the parameters can be significantly reduced taking into account the boundary conditions [24].

To parameterize photo-neutron emission and identify general prediction patterns we will act in a similar way. For this we use the results of $v(A)$ calculation for the photofission $^{235}$U and $^{238}$U at bremsstrahlung boundary energies of 12 ÷ 20 MeV [14,15].

However, prompt neutrons emission curves obtained from the calculations in [14,15] is not precise enough, so there is no sense to try to describe them adequately using complicated function with more than 2 x 2 segments.

Here, using the same methodology we simulate the behavior of neutrons from photofission of $^{235}$U and $^{238}$U actinides depending on the energy and nucleon composition in the giant dipole resonance energy range. Let us simplify this task and divide the mass interval of light and heavy fragments for the function $R(A)$ into two segments. The number of unknown parameters is also reduced while keeping the model efficiency. As a result, we get the following picture (Fig. 1).

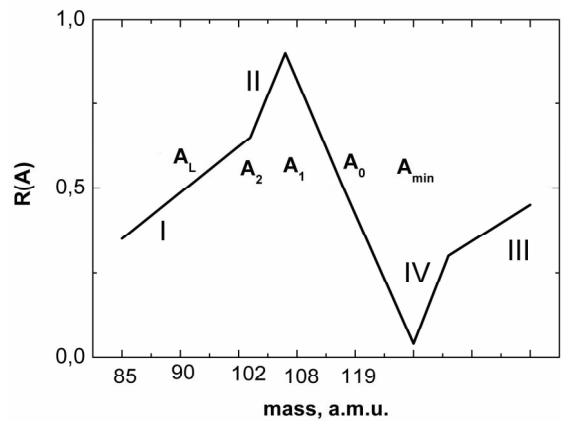

Fig. 1. An example of $R(A)$ function. The segments I-II and III-IV correspond to our parameterization.

Let us consider some features of *R(A)* function. At the point of symmetric fission $A_0 = A_F/2$; $R(A_0) = 0.5$ and $R(A_L) = 0.5$, where $A_L$ is determined from fitting.

Kink points $A_{min}$, $A_1$ and $A_2$ are chosen from physical considerations and experiment: $A_{min} = 130$ corresponds to the mass of nearly magic nucleus fragment associated with spherical shells $Z = 50$ and $N = 82$, where fission neutron yield is minimal [24]. Then maximum neutron fission yield for light fragments will match point $A_1$, which is symmetrical to $A_{min}$ relative to $A_0$, $A_1 = 2A_0 - A_{min}$ [24].

The kink point $A_2$ corresponds to the average fragment mass of light fragments, $A_2 = \langle A_L \rangle = A_F - \langle A_H \rangle$, where $\langle A_H \rangle = 138$ [27-29].

The parameters $a_i$, $b_i$ and $A_L$ are determined by calculating the function *R(A)* for 4 segments I - IV (see Fig.1). The number of free parameters can be reduced using the conditions

$$a_1 = R(A_L) = 0.5, \qquad (6)$$

$$a_2 = a_1 + (b_1 - b_2)(A_2 - A_L). \qquad (7)$$

The dependence of $b_i$ slopes on excitation energy ($E_\gamma$) is noticed on the figure for *R(A)*, so to study this dependence we have chosen $b_i = x_i + y \cdot E_\gamma$. Thus, we need to determine $A_L$, $b_i$, $x_i$ and $y$ parameters for several hundred "experimental" points.

The value of $A_L$ varies significantly with changes of actinide mass, at least for neutron-induced actinide fission (see Fig. 4) [24]. Therefore, we chose a similar parameterization [29]:

$$A_L = -80 + 1.45 A_0. \qquad (8)$$

To take even-even and even-odd effect into account we introduce the factor

$$P(N_F) = 2 - c\left[(-1)^{N_F} - (-1)^{Z_F}\right], \qquad (9)$$

$$N_F = A_F - Z_F.$$

As a result $b_i$ slopes will look as

$$b_i = (x_i + yE_\gamma) P(N_F), \qquad i = 1,2. \qquad (10)$$

We calculate the function *R(A)* for the photofission of actinides $^{235}$U and $^{238}$U according to (1) - (10) by fitting of 356 "experimental" values of *R(A)*.

Using the least squares method [30] the four parameters $x_1$, $x_2$, $c$ and $y$ were defined to satisfactorily describe the characteristic "saw-tooth" behavior of prompt neutrons from the photofission of actinides with A = 235-238 a.m.u.

Table 1. Calculated parameters of $R(A)$ function.

| Parameter | Value | Error |
|---|---|---|
| $x_1$ | $1.56 \cdot 10^{-2}$ | $0.34 \cdot 10^{-2}$ |
| $x_2$ | $0.931 \cdot 10^{-2}$ | $0.281 \cdot 10^{-2}$ |
| $y$ | $-0.315 \cdot 10^{-3}$ | $0.281 \cdot 10^{-3}$ |
| $c$ | $-0.507$ | $0.195$ |

The results of $R(A)$ calculation and prediction at the bremsstrahlung maximum energy 12 MeV are shown in Fig. 2.

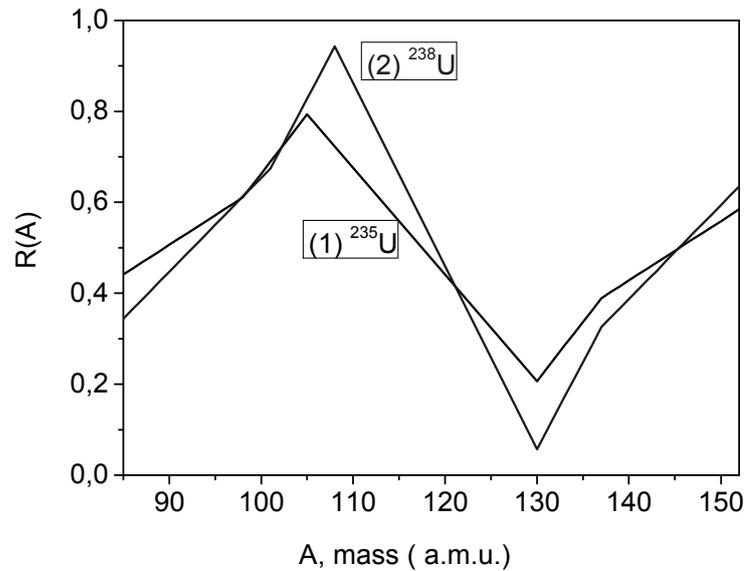

Figure 2. The result of $R(A)$ functions calculation for $^{235}$U (1), $^{238}$U (2) photofission at the bremsstrahlung maximum energy 12 MeV.

The curves for prompt neutrons yield $v_{L,H}(A)$ can be calculated with help (1). For this we need the value of the prompt neutrons averaged number $\bar{v}$ for photofission of the actinides as the function of excitation energy, mass and charge of the nuclei. One may use the results of empirical calculations of $\bar{v}$, described below.

We take into account the wide data set of photofission $\bar{v}$ $(A,Z,E_\gamma)$ in the energy range 5 – 20 MeV [32] for $^{233}$U [33, 34], $^{234}$U [33, 34], $^{235}$U [33, 35], $^{236}$U [33, 35], $^{238}$U [33, 35], $^{237}$Np [33, 34]

and $^{239}$Pu [33,34], where the approach can be applied. We selected all 219 experimental points for our calculation.

In previous studies [26, 36-39] it has been shown that expanding the charge, mass and energy dependence of $\bar{\nu}$ in the form of truncated Taylor series apparently yields a reasonable representation for $\bar{\nu}(A,Z,E_n)$ if the zero-, first- and one second-order cross term are kept in truncation for all isotopes, i. e. it is sufficient to use a linear approximation over all three variables, at least up to the threshold $(\gamma,nf)$, taking into account the contribution of the even-odd effect in a common form. As the initial formula, we have chosen:

$$\bar{\nu}(A,Z,E_\gamma) = \bar{\nu}_0(A,Z) + a(A,Z)(E_\gamma - E_s), \tag{11}$$

where the slope $\bar{\nu}_0(A, Z)$ and the intercept $a(A,Z)$ are :

$$\bar{\nu}_0(A,Z) = C_1 + C_2(Z - Z_0) + C_3(A - A_0) + C_4 P(A,Z), \tag{12}$$

and

$$a(A,Z) = C_5 + C_6(Z - Z_0) + C_7(A - A_0) + C_8 P(A,Z), \tag{13}$$

$$P(A, Z) = 2 - (-1)^{A-Z} - (-1)^Z, \tag{14}$$

$E_s$ – nucleon separation energy (see Tab.1.).

Table 1. Nucleon separation energy [40].

| Isotope | $^{232}$Th | $^{233}$U | $^{234}$U | $^{235}$U | $^{236}$U | $^{238}$U | $^{237}$Np | $^{238}$Pu | $^{239}$Pu | $^{240}$Pu | $^{241}$Pu |
|---|---|---|---|---|---|---|---|---|---|---|---|
| $E_s$ MeV | 6,4381 | 5.760 | 6,8437 | 5,2978 | 6,5448 | 6,1520 | 6,580 | 7,0005 | 5,6465 | 6,5335 | 5,2416 |

$Z_0$, $A_0$ and $E_s$ are the values, about which the expansion is to be made [36].

Coefficients $C_i$, were calculated by the least-square method [30]. The final formula for calculating the averaged number of prompt neutrons for photofission of actinides is:

$$\bar{\nu}_0(A,Z) = (1{,}97 \pm 0{,}05) + (0{,}165 \pm 0{,}028)(Z - 90) + \\ + (0{,}0341 \pm 0{,}0093)(A - 232) - (0{,}0853 \pm 0{,}0094) \cdot P(A,Z) \tag{15}$$

$$a(A, Z) = (0{,}0963 \pm 0{,}75 \cdot 10^{-2}) + (0{,}0371 \pm 0{,}43 \cdot 10^{-2})(Z - 90) - \\ - (0{,}566 \pm 0{,}138) \cdot 10^{-2} \cdot (A - 232). \tag{16}$$

The results of the $\nu_{L,H}(A)$ calculation using (1)-(11), (15), (16) and Table 1 are shown in Fig. 3 (solid curve). As can be seen from the figures the calculated values for prompt neutrons yield are everywhere within the errors.

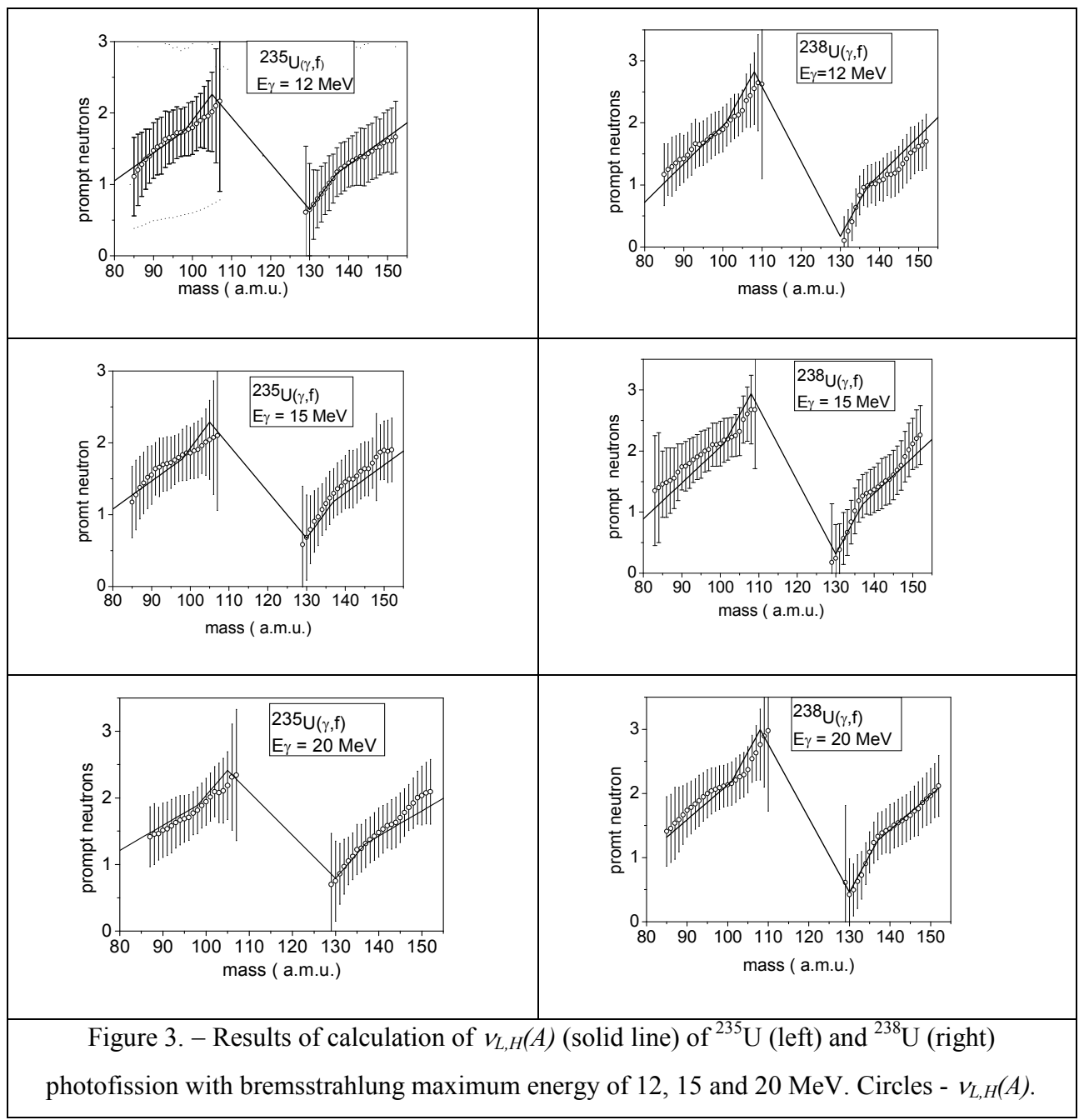

Figure 3. – Results of calculation of $\nu_{L,H}(A)$ (solid line) of $^{235}$U (left) and $^{238}$U (right) photofission with bremsstrahlung maximum energy of 12, 15 and 20 MeV. Circles - $\nu_{L,H}(A)$.

These observations allow to estimate the possible values of fission neutrons yield from light and heavy fragments with known total yields just through the mass distributions of fission fragments using the modified Terrell method [13].

### 3. CONCLUSION

With $R(A)$ function parameterization, which fairly well reproduces the characteristic features of its behavior and parameterization of averaged number of prompt neutrons, we can calculate the expected values of prompt neutrons yield for arbitrary neighboring actinides, such as $^{237}$Np or $^{239}$Pu.

Thus, to determine the photofission yield for neutron yield on fragment mass *v(A)* of arbitrary actinides we need to know the value $v_F(A)$, which is determined by the general empirical formulas (11), (15-16).

**REFERENCES**

1. *Nifenecker H., David S., Loiseau J.M., Meplan O.* Basics of accelerator driven subcritical reactors // Nuclear Instruments and Methods in Physics Research A. 2001, V. 463, Is. 3, p. 428-467.

2. *Runkle R.C., Chichester D.L., Thompson S.J.* Rattling nucleons: New developments in active interrogation of special nuclear material // Nuclear Instruments and Methods in Physics Research A. 2012, V. 663, Is 1, p. 75-79.

3. *Thompson S. J., Kinlaw M. T., Harmon J. F., et al.* Utilization of high-energy neutrons for the detection of fissionable materials // Applied Physics Letters. 2007, V. 90, Is. 7, 074106.

4. *Stevenson J., Gozani T., Elsalim M., Condron C., Brown C.* Linac based photofission inspection system employing novel detection concepts // Nuclear Instruments and Methods in Physics Research A. 2011, V. 652, Is 1, p. 124-128.

5. *Sajo-Bohus L., Greaves E.D., Davila J., Barros H., Pino F., Barrera M.T., Farina F.* Th and U fuel photofission study by NTD for AD-MSR subcritical assembly // AIP Conference Proceedings. 2015, V. 1671, 020009.

6. *Talou P., Kawano T., Chadwick M.B., Neudecker D. and Rising M.E.* Uncertainties in nuclear fission data // J. Phys. G: Nucl. Part. Phys. 2015, V. 42, 034025 (17pp)

7. *Silano J.A.* Sub-barrier photofission measurements in $^{238}$U and $^{232}$Th // A dissertation submitted to the faculty of the University of North Carolina at Chapel Hill in partial fulfillment of the requirements for the degree of Doctor of Philosophy in the Department of Physics and Astronomy. Chapel Hill 2016. 164 p.
https://cdr.lib.unc.edu/indexablecontent/uuid:b6b4f9a1-eb77-4214-a7ba-98700c2eb141

8. *Wahl A.C., Ferguson R.L., Nethaway D.R.* Nuclear-Charge distribution in low-energy fission. // Physical Review. 1962, V. 126, Is. 3, p. 1112-1127.

9. *Naik H., Dange S.P., Reddy A.V.R.* Charge distribution studies in the odd-Z fissioning systems // Nuclear Physics A. 2007, V. 781, Is. 1-2, p. 1-25.

10. *M.N. bin Mohd Nasir, Metorima K., Ohsawa T., Hashimoto K.* Analysis of incident-energy dependence of delayed neutron yields in actinides// AIP Conference Proceedings. 2015, V. 1659, 040002-1-7

11. *Naik H., Nimje V.T., Raj D. et al.* Mass distribution in the bremsstrahlung-induced fission of $^{232}$Th, $^{238}$U and $^{240}$Pu // Nuclear Physics A. 2011, V. 853, Is. 1, p. 1-25.

12. *Brose U.* Sawtooth curve of neutron multiplicity // Physical Review C. 1985, V. 32, Is. 4, p. 1438-1441.

13. *Terrell J.* Neutron Yields from Individual Fission Fragments // Physical Review. 1962, V. 127, Is 3, p. 880-904.

14. *Piessens M., Jacobs E., De Frenne D. et al.* Photon induced fission of $^{232}$Th with 12 and 20 MeV bremsstrahlung // Proceedings of the XV-th International Symposium on Nuclear Physics (Nuclear Fission). November 11 - 15, 1985 in Gaussig. 1986, p. 92-95.
https://www-nds.iaea.org/publications/indc/indc-gdr-0042G.pdf

15. *De Frenne D., Thierens H., Proot B. et al.* Charge distribution for photofission of $^{235}$U and $^{238}$U with 12-30 MeV bremsstrahlung // Physical Review. C. 1982, V. 26, Is. 4, p. 1356-1368.

16. *Jacobs E., Thierens H., De Clercq A. et al.* Neutron emission in the photofission of $^{235}$U and $^{238}$U with 25-MeV bremsstrahlung // Physical Review. C. 1976, V. 14, Is. 5, p. 1874-1877.

17. *Jacobs E., De Clercq A., Thierens H. et al.* Fragment mass and kinetic energy distributions for the photofission of $^{238}$U with 12-, 15-, 20-, 30-, and 70-MeV bremsstrahlung // Physical Review C. 1979, V. 20, Is. 6, p. 2249-2256.

18. *Jacobs E., De Clercq A., Thierens H. et al.* Fragment mass and kinetic energy distributions for the photofission of $^{235}$U with 12-, 15-, 20-, 30-, and 70-MeV bremsstrahlung // Physical Review C. 1981, V. 24, Is. 4, p. 1795-1798.

19. *Jacobs E., Thierens H., De Frenne D. et al.* Product yields for the photofission of $^{235}$U with 12-, 15-, 20-, 30-, and 70-MeV bremsstrahlung // Physical Review C. 1980, V. 21, p. 237-245.

20. *Jacobs E., Thierens H., De Frenne D. et al.* Product yields for the photofission of $^{238}$U with 12-, 15-, 20-, 30-, and 70-MeV bremsstrahlung // Physical Review C. 1979, V. 19, p. 422-432.

21. *Wang F. and Hu J.* Transformation between pre- and post-neutron-emission fragment mass distribution // in China Nuc. Sci. Tech. Report, 1989, CNIC-00293.

22. *Tan Jiawei, Bendahan Joseph* Geant 4 modifications for accurate fission simulations // Physics Procedia. 2017, V. 90, p. 256-265.

23. *Lengyel A.I., Parlag O.O., Maslyuk V.T., Kibkalo Yu.V.* Fission neutron yields from product mass distribution // Proceedings the 3-rd International Conference "Current problems in nuclear physics and atomic energy" Kyiv 2011. p. 476-478.
http://www.kinr.kiev.ua/NPAE-Kyiv2010/html/Proceedings/5/Lengyel.pdf

24. *Lengyel A.I., Parlag O.O., Maslyuk V.T., Kibkalo Yu.V.* Phenomenological description of neutron yields from actinide fission // Problems of atomic science and technology, 2011, N3. Series: Nuclear physics investigations (55), p.14-18.
http://vant.kipt.kharkov.ua/ARTICLE/VANT_2011_3/article_2011_3_14.pdf

25. *Wahl A.C.* // Nuclear-charge distribution and delayed-neutron yields for thermal-neutron-induced fission of $^{235}$U, $^{233}$U, and $^{239}$Pu and for spontaneous fission of $^{252}$Cf // Atomic Data and Nuclear Data Tables. 1988, V. 38, Is. 1, p. 1-156.

26. *Wahl A.C.* Systematics of fission product yields // Fission product yield data for the transmutation of minor actinide nuclear waste//IAEA 2008, 117-148.
http://www-pub.iaea.org/MTCD/publications/PDF/Pub1286_web.pdf

27. *Al-Adili A., Hambsch F.-J., Pomp S., Oberstedt S.* Impact of prompt-neutron corrections on final fission-fragment distributions // Physical Review C. 2012, V. 86, Is. 5. 054601[8 pages]

28. *Rubchenya V.A., Äystö J.* Consistent theoretical model for the description of the neutron-rich fission product yields // The European Physical Journal A. 2012, V. 48, Is. 1, p. 1-8.

29. *Lengyel A.I., Parlag O.O., Maslyuk V.T., Kibkalo Yu.V., Romanyuk M.I.* Parametrisation of prompt neutron yields from photofission fragments of actinide nuclei for the giant dipole resonance energy range // Problems of atomic science and technology. 2014, №5 (95), Series Nuclear Physics Investigations (63), p. 12-17. http://vant.kipt.kharkov.ua/ARTICLE/VANT_2014_5/article_2014_5_12.pdf

30. *James F., Ross M.* Function minimization and error analysis. MINUIT D506. CEREN Computer Centre Program library. 1967, p. 1-47.

31. *Nishinaka I., Tanikawa M., Nagame Y., Nakahara H.* Nuclear-charge polarization at scission in proton-induced fission of $^{233}$U // The European Physical Journal A. 2011, V. 47, p. 1-8.

32. *Lengyel A.I., Parlag O.O., Maslyuk V.T., Romanyuk M.I., Gritzay O.O.* Calculation of average numbers of prompt neutrons for actinide photofission // Journal of Nuclear and Particle Physics. 2016, V. 6(2), p. 43-46. (DOI: 10.5923/j.jnpp.20160602.03)
http://article.sapub.org/10.5923.j.jnpp.20160602.03.html

33. *Caldwell J.T., Dowdy E.J.* Experimental determination of photofission neutron multiplicities for eight isotopes in the mass range 232 ≤ A ≤ 239 // Nuclear Science and Engineering. 1975, V. 56, p. 179-187.

34. *Berman B.L., Caldwell J.T., Dowby E.I. et al.* Photofission and photoneutron cross sections and photofission neutron multiplicities for $^{233}$U, $^{234}$U, $^{237}$Np and $^{239}$Pu // Physical Review C. 1986, V. 34, Is. 6, p. 2201-2214.

35. *Caldwell J.T., Alvarez R.H., Berman B.L. et al.* Experimental determination of photofission neutron multiplicities for $^{235}$U, $^{236}$U, $^{238}$U, and $^{232}$Th using monoenergetic photon // Nuclear Science and Engineering. 1980, V. 73, N 2, p. 153-163.

36. *Howerton R.J.* $\bar{\nu}$ revised // Nuclear Science and Engineering. 1977, V. 62, p. 438-454.

37. *Bois R., Frehaut J.* Evaluation semi-empirique de $\bar{\nu}_P$ pour la fission induite par neutrons rapides // Commissariats a l'Energie Atomique report CEA-R-4791. (1976). p. 44-51.
http://www.iaea.org/inis/collection/NCLCollectionStore/_Public/08/297/8297945.pdf

38. *Ohsawa T.* Empirical formulas for estimation of fission prompt neutron multiplicity for actinide nuclides // Journal of Nuclear and Radiochemical Sciences. 2008, V. 9, No.1, p. 19-25.

39. *Lengyel A.I., Kibkalo Yu.V., Parlag O.O., Maslyuk V.T.* Calculations of the yield of prompt neutrons from fission fragments // Uzhhorod university scientific herald. Series Physics. 2008, N23, p.58-63 (in Ukrainian). http://nbuv.gov.ua/j-pdf/Nvuufiz_2008_23_9.pdf

40. *Verbeke J.M., Hagmann C., Wright D.* Simulation of neutron and gamma ray emission from fission and photofission. LLNL Fission Library 2.0.2. // UCRL-AR-228518-REV-1. Lawrence Livermore National Laboratory. October 24, 2016.
https://nuclear.llnl.gov/simulation/fission.pdf